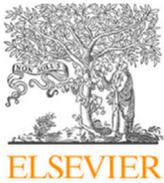
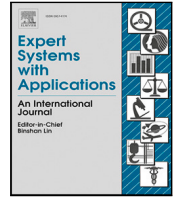

# FGR-Net: Interpretable fundus image gradeability classification based on deep reconstruction learning

Saif Khalid [a,b,*], Hatem A. Rashwan [a], Saddam Abdulwahab [a], Mohamed Abdel-Nasser [a,c], Facundo Manuel Quiroga [d], Domenec Puig [a]

[a] *Department of Computer Engineering and Mathematics, Universitat Rovira i Virgili, 43307 Tarragona, Spain*
[b] *University of Al-Qadisiyah, 58002 Al Diwaniyah, Iraq*
[c] *Department of Electrical Engineering, Aswan University, 81528 Aswan, Egypt*
[d] *Instituto de Investigación en Informática LIDI, Facultad de Informática, Universidad Nacional de La Plata, 7776, B1900 La Plata, Provincia de Buenos Aires, Argentina*

ABSTRACT

The performance of diagnostic Computer-Aided Design (CAD) systems for retinal diseases depends on the quality of the retinal images being screened. Thus, many studies have been developed to evaluate and assess the quality of such retinal images. However, most of them did not investigate the relationship between the accuracy of the developed models and the quality of the visualization of interpretability methods for distinguishing between gradable and non-gradable retinal images. Consequently, this paper presents a novel framework called "FGR-Net" to automatically assess and interpret underlying fundus image quality by merging an autoencoder network with a classifier network. The FGR-Net model also provides an interpretable quality assessment through visualizations. In particular, FGR-Net uses a deep autoencoder to reconstruct the input image in order to extract the visual characteristics of the input fundus images based on self-supervised learning. The extracted features by the autoencoder are then fed into a deep classifier network to distinguish between gradable and ungradable fundus images. FGR-Net is evaluated with different interpretability methods, which indicates that the autoencoder is a key factor in forcing the classifier to focus on the relevant structures of the fundus images, such as the fovea, optic disk, and prominent blood vessels. Additionally, the interpretability methods can provide visual feedback for ophthalmologists to understand how our model evaluates the quality of fundus images. The experimental results showed the superiority of FGR-Net over the state-of-the-art quality assessment methods, with an accuracy of > 89% and an F1-score of > 87%. The code is publicly available at https://github.com/saifalkh/FGR-Net.

## 1. Introduction

Fundus retinal photography uses a fundus camera to record color images of the eye's internal surface condition to screen for eye disorders and track their progression. Various eye disorders, such as diabetic retinopathy (DR) (Mookiah et al., 2013), cataract (Guo, Yang, Peng, Li, & Liang, 2015), age-related macular degeneration (AMD) (Akram, Tariq, Khan, & Javed, 2014), and glaucoma (Joshi, Sivaswamy, & Krishnadas, 2011) are diagnosed using fundus imaging. These diseases affect a considerable percentage of the world's population. However, while ophthalmologists strive to provide appropriate medical care to many patients, the number of eye specialists available to satisfy the current demand is insufficient (International Council of Ophthalmology, 2019).

Artificial Intelligence (AI) has recently played a significant role in capturing, evaluating, and analyzing fundus images. AI-based fundus image analysis systems help reduce the shortage of ophthalmologists by providing accurate and quick diagnoses of thousands of fundus images. Many AI-powered approaches for screening and diagnosing various eye diseases have been proposed in the literature (Baget-Bernaldiz et al., 2021; Gong, Kras, & Miller, 2021; Keenan et al., 2022; Ramachandran, Hong, Sime, & Wilson, 2018). However, low-quality fundus images degrade the performance of AI-based fundus image analysis systems (Pérez, Perdomo, & González, 2020). Ophthalmologists have criteria to grade the quality of retinal images before treatment and diagnosis. In the case of fundus images, this process is called Image Quality

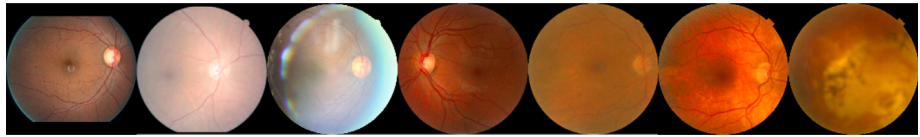

**Fig. 1.** Samples from three databases: EyeQuality (Good and usable and reject) (Fu et al., 2019), EyePACS (gradable, ungradable) (Foundation, 2019), and our private dataset (gradable, ungradable) respectively for retinal fundus images.

Assessment (IQA) in general or Image Gradability Classification. The process determines whether an image can be used for diagnosis.

Fig. 1 presents examples of fundus images acquired by retinography devices. These images suffer from different distortions, such as color distortion, uneven illumination, and low contrast. Reliable screening of eye diseases requires fundus images of sufficient quality to analyze and extract disease biomarkers. Therefore, various fundus image gradability approaches have been proposed to check the gradability of the images before feeding them into the diagnosis systems. Different fundus image gradability methods have been presented in the literature based on handcrafted computer vision-based techniques to classify input images as gradable or ungradable (Wang et al., 2015; Yu et al., 2012; Zheng et al., 2010). Such handcrafted fundus image assessment techniques employ morphological feature representations to detect the local anatomical properties. However, the image features generated are *problem-specific* and *rarely reusable*, i.e., not guaranteed to work for other images acquired by different fundus cameras.

Medical imaging analysis based on deep neural networks (DNNs) has witnessed advanced progress during the last decade, such as Dash et al. (2022), Wieczorek, Siłka, Woźniak, Garg, and Hassan (2021) and Woźniak, Siłka, and Wieczorek (2021). With the era of deep learning, many such architectures have been employed to build fundus image gradability models (Costa et al., 2017; Pérez et al., 2020). For instance, Pérez et al. (2020) presented a deep learning-based model for assessing fundus image quality, a lightweight model executed on different devices. Most deep learning-based image quality assessment methods use classification models based on Convolutional Neural Networks (CNN). However, the performance of these gradability methods significantly degrades with fundus images of various fundus cameras since CNNs can only generate representations similar to the training data. Moreover, the encoder network can lose much information about the input image by increasing the built-in convolutional layers, and the classifier layer can yield misclassified results. *To handle this issue*, in this paper, we propose an autoencoder network to learn a compressed input feature model before classifying the fundus images' quality. Such features are generated because the autoencoder is forced to prioritize which characteristics of the input image should be encoded, learning valuable properties from the data. An autoencoder is a self-supervised deep learning network comprising an encoder and a decoder sub-networks. The encoder compresses the input, and the decoder attempts to reconstruct the input from the compressed version provided by the encoder. After the training stage of the FGR-Net model, the encoder and the classifier models are saved, and the decoder is discarded from the model. In the testing phase, the encoder can then be used as a feature extraction technique on the input data, using the features as input for the classifier network. Fig. 2 shows the framework of the proposed image gradability classification method "FGR-Net".

It should be noted that most existing deep learning-based fundus image gradability methods are black boxes. The black-box nature of the developed models causes distrust in ophthalmologists, thus reducing the use of AI-based solutions in clinical practice. To handle this issue, in this paper, we propose using *different interpretability techniques* such as saliency maps (Arun et al., 2021) in the fundus image quality assessment framework. Such techniques can reveal the disturbed and abnormal regions in the fundus images to explain why the model provides gradability classification results.

This work extends our preliminary work presented in Khalid et al. (2021). The most significant contributions of this work are:

- A precise deep learning-based fundus image gradability model consists of two parallel networks. The first is an autoencoder network composed of two successive networks; encoder and decoder. The encoder will extract the fundus images' representative anatomical characteristics and structures properly. The decoder reconstructs the input image from the feature extracted by the encoder network based on self-supervised learning by comparing the reconstructed image to the input image. The second is a CNN-based fundus image gradability classifier fed by the features learned by the encoder network to classify input fundus images as gradable or ungradable.
- Different off-the-shelf interpretability methods are integrated with the FGR-Net model to generate interpretable visual feedback for ophthalmologists to understand why our model grades the quality of fundus images. Our interpretability analysis indicates that the autoencoder network helps the classifier to focus more on the relevant structures of the fundus images, such as the fovea, optic disk, and prominent blood vessels. On the other hand, the CNN-based classifier model uses more arbitrary input regions to determine the gradability of the image, which is often not relevant anatomically.

The remainder of the paper is organized in the following manner. Section 2 discusses related works on retinal image gradability assessment and model interpretability. The FGR-Net model is presented in Section 3. Section 4 explains how this paper enables the Explainability of FGR-Net. Section 5 provides and discusses the experimental results with three datasets. Section 6 concludes the article and gives directions for the future scope of our research.

## 2. Related work

### 2.1. Retinal image gradability assessment

Many fundus image gradability methods are based on two-class quality labels (i.e., 'Accept' and 'Reject'). Others are based on a three-class quality grading system (i.e., 'Good', 'Usable', and 'Reject'). For instance, Yu et al. (2012) proposed a method for automatically classifying the quality of retinal images based on the RGB color space. Their system uses vessel density, textural features, global histogram features, and a metric known as non-reference perceptual sharpness. They also concentrated on three Regions of Interest (ROI); lower retinal hemispheres, upper retinal hemispheres, and optic disk regions. Wang et al. (2015) introduced a retinal image assessment algorithm that selects images of acceptable generic quality. The algorithm uses three human visual system characteristics: multi-channel sensation, noticeable blur, and contrast sensitivity functions to detect illumination and color distortion, blur, and low contrast distortion. They used a total of 536 retinal images, 280 from proprietary datasets, and 256 from public databases (DRIMDB (Prentašić et al., 2013), and DRIVE (Staal, Abràmoff, Niemeijer, Viergever, & Van Ginneken, 2004)). Then, they employed Support Vector Machines (SVMs)and a Decision Tree for binary classification. They achieved a sensitivity of 87.45% and a specificity of 91.66%. A recent work, Karlsson et al. (2021), proposed a method that automatically grades image quality on a continuous scale. The technique utilizes random forest regression models trained on image features discovered automatically using Fourier transform. The method was tested on DRIMDB, a publicly available dataset with binary quality ratings. They used 194 fundus images annotated into 125 good and 69 bad-quality images annotated by a medical expert. Their method achieves an accuracy of 98%, a sensitivity of 99%, and a specificity of 95%. In turn, Avilés-Rodríguez et al. (2021) introduced a method called topological data analysis (TDA), for image quality assessment of eye fundus images based on extracting topological descriptors integrated into a machine learning classifier. They used the EyePACS dataset as a binary classification between images with quality (Good) and (Bad) by randomly selecting subsets of 2000 images for each class. Avilés-Rodríguez et al. (2021) showed a average precision of 93%, and AUC of 98%. However, these results were achieved with a very small set of fundus images.

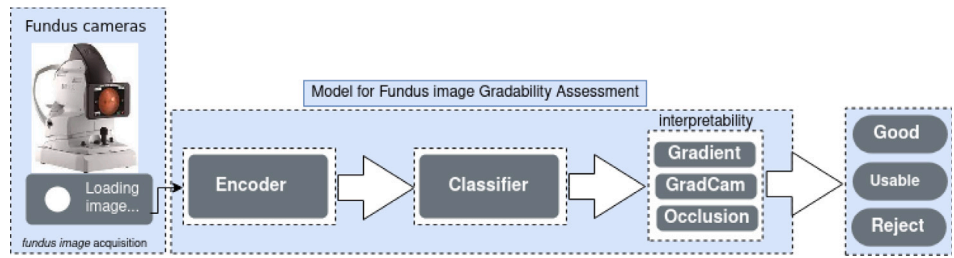

**Fig. 2.** An overview of the FGR-Net model. A retinography device generates a fundus image, which our model processes. The model outputs gradability results and interpretability feedback to ophthalmologists.

In addition, Recently, Raj, Shah, Tiwari, and Martini (2020) suggested a multivariate regression-based CNN model for fundus image quality assessment. They evaluated their proposed system on the FIQuA dataset (Raj et al., 2020) that contains 1500 fundus images (Foundation, 2019) and obtained a classification rate of 95.66%.

In turn, Xu, Liu et al. (2020) proposed SalStructuIQA that mimics how ophthalmologists assess the quality of retinal images. They extracted two types of salient structures from fundus retinal images: large structures, such as the optic disk, and small structures, mainly vessels. The SalStructuIQA model incorporates the proposed two salient structure priors with a deep CNN model to classify the quality of the input images. SalStructuIQA was evaluated on the public Eye-Quality dataset (Fu et al., 2019). SalStructuIQA achieved Precision and F1-scores of 87% and 87%, respectively. Fu et al. (2019) analyzed the effects of different color spaces on retinal image evaluation and proposed a deep network called Fusion Color-space Network (MCF-Net). To predict image quality scores, MCF-Net combined other color space representations (i.e., RGB, HSV, and LAB) at the feature and prediction levels. MCF-Net was assessed on the public Eye-Quality dataset (Fu et al., 2019) public Eye-Quality dataset (Fu et al., 2019). MCF-Net achieved Precision and F1-score of 86% and 85%, respectively. Nderitu et al. (2021) introduced a deep learning model based on the Efficient-Net-B0 network to classify retinal image gradability effectively. Their model provided Area Under the Curve (AUC) of 0.93 for binary classification following five-fold cross-validation and a Kappa coefficient of 0.69. Muddamsetty and Moeslund (2020) proposed the RFIQA task for multi-level retinal fundus image quality assessment. RFIQA predicts six quality grades based on essential regions of the fundus images for diagnosing diabetic retinopathy (DR), AgedMacular Degeneration (AMD), and Glaucoma. RFIQA contains 9945 fundus images captured by different fundus cameras and under various imaging conditions from multiple patients with other retinal diseases. The dataset has six levels of quality as grades Grade 0 (Good) of 5444 images, Grade 1 (Good; Periphery Not Visible) of 1817 images, Grade 2 (Bad; Optical Disk Not Visible) of 158 images, Grade 3 (Bad; Macula Area Not Clearly Visible) of 1058 images, Grade 4 (Bad; Unsharp, Blinking, Big Reflections, Over Exposure) of 1449 images, and Grade 5 (Bad; Miscellaneous) of 19 images. The authors combined a deep CNN model and generic texture features to extract patterns of the images and then used a Random Forest algorithm as a classifier. They achieved an F1-score of 87%. Furthermore, Dai, Wu et al. (2021) suggested a deep learning model composed of three sub-networks: image quality assessment sub-network, lesion-aware subnetwork, and DR grading sub-network. They used a dataset of 5176 retinal images initially captured from 1294 patients. 1487 images of these images were recognized as low-quality with artifacts, clarity, and field definition issues and 3689 images were of adequate quality. The dataset is used for training the image quality assessment sub-network. Their proposed model provided classification rate comparable to the state-of-the-art.

The above methods have solved the quality assessment problem as a classification problem. Besides, the classification rate is still not enough to use these developed models in a clinical setup. In addition, most retinal image quality assessment models lack interpretability tools for the proposed networks. Thus, this paper focuses both on improving the performance of automatic retinal quality assessment models and providing an interpretable assessment model. To the best of our knowledge, no method has been proposed to classify the quality of retinal images through a model merging between an autoencoder network based on self-supervision for image reconstruction and a supervised classifier that helps in enhancing the feature extraction process of the input fundus images.

### 2.2. Interpretability for fundus analysis models

Explainable machine/deep learning models help clinicians interpret black-box models and their decision-making process to verify why these models took a particular decision, which is extremely important in the medical field. Explainability is a desired feature in models that allows both users and researchers to understand relationships between model inputs, outputs and domain concepts (Escalante et al., 2018). These can be classified into three broad groups (Escalante et al., 2018):

- Rule-extraction methods, which infer high-level rules from the relationship between inputs and outputs of a network.
- Attribution methods, which measure the importance of a *component* by changing to the input or internal components and recording how much the changes affect model performance. Attribution methods are often visualized and sometimes referred to as visualization or saliency methods.
- Intrinsic methods, which aim to improve the interpretability of internal representations with methods that are part of the model architecture. Intrinsic methods increase fidelity, clarity, and parsimony in attribution methods.

This work focuses on interpretability methods for deep learning models in computer vision. In particular, for computer vision techniques based on deep learning models, most interpretability methods visualize the information obtained by attribution methods. Visualization methods were popularized by Erhan, Bengio, Courville, and Vincent (2009), Simonyan, Vedaldi, and Zisserman (2013) and Zeiler and Fergus (2014) in recent years and provide various ways to visualize important features of a model. These are intuitive methods to gain various insights about a deep neural networks (DNN) decision process on many levels, including architecture assessment, model quality assessment and even user feedback integration.

The most basic visualization method consists of visualizing the feature maps or *activations* of the network for individual inputs (Zeiler & Fergus, 2014). Given the large number of feature maps in a typical CNN network, the outputs of this technique can be cumbersome to analyze (Cammarata et al., 2020). Hence, other attribution methods have been developed to directly relate the output layers' scores to the values of the input layer.

Recently, researchers have been increasingly using interpretability approaches to understand model decisions for medical diagnosis in various problems, including fundus image models, which are summarized below:

- In Xu, Zou, and Liu (2020, 2021), the authors included priors for brightness and large-size and tiny-size structures to enable the visualization of the importance of features based on these types of structures and brightness regions separately. Afterwards, they use GradCAM to verify that the model learns particular features corresponding to each prior.
- Jang, Son, Park, Park, and Jung (2018) used GradCAM (Selvaraju et al., 2017) to compare the representations of a simple neural network with VGG-16 and AlexNet for Laterality Classification. For a few examples, they verify that the network focuses on features around the optic disk and prominent blood vessels.
- In Jiang et al. (2019), an ensemble model is proposed for Diabetic Retinopathy diagnosis, and GradCAM was used to compare the representation between vanilla neural network GradCAM and the ensemble, but no conclusions could be drawn from the comparison.
- In de La Torre, Valls, and Puig (2020), saliency activations were employed to visualize the feature maps of specific layers of the model with a small set of samples.

While most existing works focus on interpreting models for diagnosis (detection or grading) with fundus images, to the best of our knowledge there are no previous works that concentrate on interpretability for gradability (or image quality). Interpreting models for estimating gradability is subtly different from diagnosis models since the image quality considers both global (illumination, texture, focus) and local (anatomical structures) characteristics of the images. Therefore, a model for gradability may use or take advantage of more obscure features of an image, such as low level texture information that is hard to spot for humans but useful for tasks. This fact complicates the interpretability of the model compared to diagnostic models.

To the best of our knowledge, Shen et al. (2020) is the only previous work that applies interpretability methods to the problem of fundus gradability. In that case, they added fovea and optical disk localization as secondary tasks for a fundus image quality classifier. Afterwards, they visually use GradCAM to visually confirm the model's understanding of the fundus image. However, they require manual annotation of these landmarks, which are seldom available and hard to define for low-quality fundus images.

In this work, we focus on interpretability models that can be used in real-time to provide feedback to medical practitioners during the capture of fundus images. As mentioned before, we propose utilizing an autoencoder network to improve the intermediate representation of the image. Given that interpretability methods such as GradCAM focus on intermediate representations, this coupling gives better visualizations in the interpretability methods.

### 2.3. Contributions

Previous and recent studies in the field of fundus image classification are an important topic because of their relationship to health care in today's society and the field of medical image processing research. Traditional machine learning methods generally produce more interpretable classification models. but they only perform better on small sample datasets. On the other hand, deep learning methods can utilize large datasets and models to achieve state-of-the-art performance. Furthermore, deep learning methods have achieved state of the art results in ophthalmology-related tasks.

In our work, we focus on solving the fundus image quality problem. We achieve better results than traditional deep learning networks by using an auxiliary autoencoder network to reconstruct the input image. The autoencoder helps to improve the intermediate representations and focus on relevant features to grade the quality of the fundus images. To validate this claim, we supplemented our model with interpretability methods to understand which features are taken into account. Additionally, the same interpretability techniques can help ophthalmologists and experts distinguish between gradable and ungradable images for timely recapture.

## 3. Methodology

This section explains the FGR-Net model for retinal image gradability classification. Besides, we demonstrate interpretability techniques to provide visual feedback for doctors to understand which landmarks FGR-Net looks for when classifying the gradability of input images.

### 3.1. FGR-Net model

Fig. 3 depicts the FGR-Net model for fundus image gradability assessment. The first (top) part of the network is an autoencoder trained to learn robust feature representations of fundus images. The intermediate representations learned by the autoencoder are fed into a classifier to predict the gradability of the input fundus images as three labels: Good, Usable, or Reject.

In FGR-Net, we present a self-supervised approach for image–image translation. To formulate the reconstruction for the fundus image, let $A \in \mathbb{A}$ be a fundus image. The problem of generating a reconstructed image, $B \in \mathbb{A}$, can be formally defined as a function: $f : \mathbb{A} \rightarrow \mathbb{A}$, that maps elements from a domain $\mathbb{A}$ to the same domain $\mathbb{A}$, under a constraint of that the representation of the input image must be encoded into a lower-dimensional manifold to force the compression of the input features. To optimize the autoencoder network, it is trained via backpropagation using as a loss function minimizing the distance between the reconstructed image and the input image (i.e., target).

The autoencoder network helps to learn the fundus image-relevant features of the input fundus image, including the visible quality features. Thus, the input to our autoencoder is an RGB fundus image, and the target is the same as the input fundus image. We propose that if the autoencoder network succeeds in reconstructing the same input image, the network succeeds in learning the input image's key features, including visual quality features. In this way, we can ensure that the intermediate representations preserve the information required for the gradability classification task. The experiments support our hypothesis. The proposed autoencoder network contains two sub-networks: encoder and decoder.

*Encoder:* The encoder that obtains different levels of abstraction of fundus image features is continuously sampled through five blocks. Each block consists of a convolutional layer with a kernel size of $3 \times 3$ and an activation function of ReLU followed by a max-pooling of $2 \times 2$. The input image size to the encoder is $480 \times 480 \times 3$, and the output is a feature map of $15 \times 15 \times 512$. The detailed structure is shown in Table 1.

*Decoder:* The decoder consists of five deconvolution layers (i.e., up-sampling using bilinear interpolation and a convolutional layer with a kernel of $3 \times 3$) on the top-level feature map extracted from the encoder network to combine different features in the downsampling process and restore the input fundus image. Skip connections were used to connect the corresponding layers between the encoder and decoder networks to preserve the spatial information and the anatomical structures in fundus images. At the top of the decoder, a convolutional layer with a kernel size of 1 × 1 is used to reconstruct the image. The detailed structure is shown in Table 2.

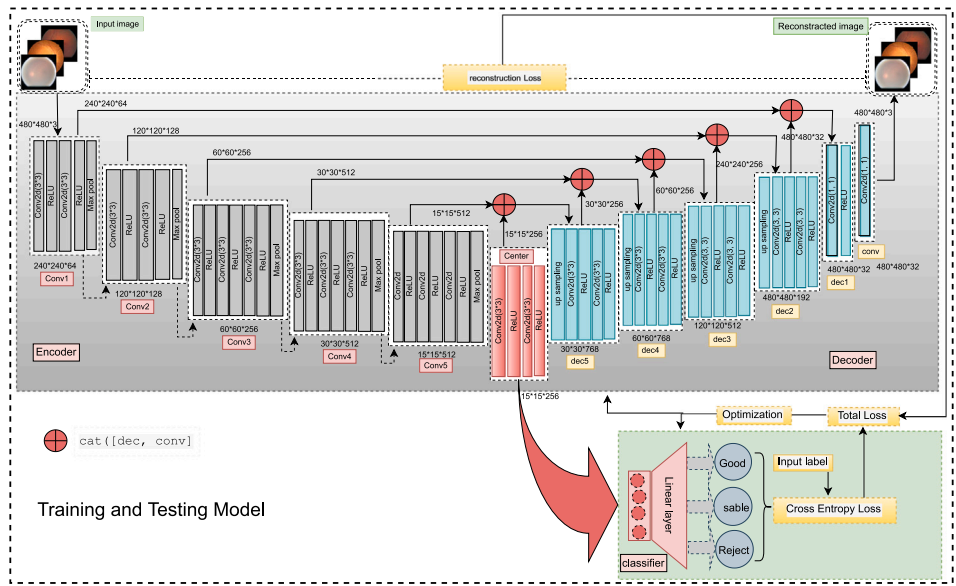

**Fig. 3.** General overview of the FGR-Net model in the train and test stages.

**Table 1**
The detailed structure of the encoder network.

|  | Input | Layer type | Filters | Kernel_size | Stride and padding | Output shape |
|---|---|---|---|---|---|---|
| (conv1) | 480 × 480 × 3 | Conv2d + ReLu | 64 | 3 | (1, 1),(1, 1) | 240 × 240 × 64 |
|  | 240 × 240 × 64 | Conv2d + ReLu | 64 | 3 | (1, 1),(1, 1) | 240 × 240 × 64 |
|  | 240 × 240 × 64 | MaxPool2d | – | 2 | 2,0 | 120 × 120 × 64 |
| (conv2) | 120 × 120 × 64 | Conv2d + ReLu | 128 | 3 | (1, 1),(1, 1) | 120 × 120 × 128 |
|  | 120 × 120 × 64 | Conv2d + ReLu | 128 | 3 | (1, 1),(1, 1) | 120 × 120 × 128 |
|  | 120 × 120 × 64 | MaxPool2d | – | 2 | 2,0 | 60 × 60 × 128 |
| (conv3) | 60 × 60 × 128 | Conv2d + ReLu | 256 | 3 | (1, 1),(1, 1) | 60 × 60 × 256 |
|  | 60 × 60 × 128 | Conv2d + ReLu | 256 | 3 | (1, 1),(1, 1) | 60 × 60 × 256 |
|  | 60 × 60 × 128 | MaxPool2d | – | 2 | 2,0 | 30 × 30 × 256 |
| (conv4) | 30 × 30 × 256 | Conv2d + ReLu | 512 | 3 | (1, 1),(1, 1) | 30 × 30 × 512 |
|  | 30 × 30 × 256 | Conv2d + ReLu | 512 | 3 | (1, 1),(1, 1) | 30 × 30 × 512 |
|  | 30 × 30 × 256 | Conv2d + ReLu | 512 | 3 | (1, 1),(1, 1) | 30 × 30 × 512 |
|  | 30 × 30 × 256 | MaxPool2d | – | 2 | 2,0 | 15 × 15 × 512 |
| (conv5) | 15 × 15 × 512 | Conv2d + ReLu | 512 | 3 | (1, 1),(1, 1) | 15 × 15 × 512 |
|  | 15 × 15 × 512 | Conv2d + ReLu | 512 | 3 | (1, 1),(1, 1) | 15 × 15 × 512 |
|  | 15 × 15 × 512 | Conv2d + ReLu | 512 | 3 | (1, 1),(1, 1) | 15 × 15 × 512 |
|  | 15 × 15 × 512 | MaxPool2d | – | 2 | 2,0 | 15 × 15 × 256 |

*Classifier:* The resulting feature map with a size of 15 × 15 × 512 of the autoencoder network is fed into a classifier network to classify the retinal fundus image quality into the corresponding gradability categories. The classifier network consists of four fully connected (FC) layers. See Table 3 for a detailed structure of the classifier network.

In the training phase, the whole network merging the autoencoder and classifier networks shown in Fig. 3 is optimized. While in the testing phase, we use only the trained encoder and classifier networks to classify the quality of fundus images into three classes: Good, Usable and Reject images.

### 3.2. Training

In this work, we tested the performance of the model with three different reconstruction loss functions, $L_{rec}$, used to compare the input images to the reconstructed images under self-supervision learning. Notably, we tested simple and standard loss functions as a reconstruction loss to support our idea that the autoencoder itself, not the loss function, helps the network find relevant patterns related to the characteristics of the image's quality.

The first tested reconstruction loss function $L_{rec}$ is to compute the mean square error (MSE) between the actual input image of input $A$ and the reconstructed image of $\hat{A}$ resulting from the Autoencoder network. MSE is the average of squared differences between the actual and expected values that can be defined as:

$$L_{rec}(\hat{A}, A) = \frac{1}{n}\sum_{i=1}^{n}(A_i - \hat{A}_i)^2, \qquad (1)$$

where $A_{(i)}$ is the input image of pixel $i$, $\hat{A}_{(i)}$ is the reconstructed image, and the $n$ is the number of pixels in the image $A$.

The second tested reconstruction loss function, $L_{rec}$, is the Mean absolute error (MAE). MAE is the mean of the absolute differences between actual and predicted values that can be defined as:

$$L_{rec}(\hat{A}, A) = \frac{1}{n}\sum_{i=1}^{n}|A_i - \hat{A}_i|. \qquad (2)$$

The third reconstruction loss function, $L_{rec}$, is a structural similarity index measure (SSIM), a method for predicting the perceived quality of digital images. SSIM is used for measuring the similarity between two images. The SSIM index is a complete reference metric for measuring the quality of reconstructed images compared to input images. Contrary to the L1 and L2 losses, the SSIM metric measures the similarity by comparing two images based on three aspects related to image quality: luminance, contrast and structural information. Thus, SSIM is a well-established metric for assessing the difference between two images. SSIM can be defined as:

$$L_{rec}(\hat{A}, A) = \frac{(2\mu_{\hat{A}}\mu_A + c_1)(2\sigma_{\hat{A}A} + c_2)}{(\mu_{\hat{A}}^2 + \mu_A^2 + c_1)(\sigma_{\hat{A}}^2 + \sigma_A^2 + c_2)}, \quad (3)$$

where $\mu_{\hat{A}}$ is the mean of $\hat{A}$, $\sigma_{\mu_{\hat{A}}}$ is the standard deviations of $\hat{A}$, $\mu_A$ is the mean of $A$, $\sigma_{\mu_A}$ is the standard deviations of $A$, $\sigma_{\hat{A}A}$ is the covariance of $\hat{A}$ and $b$, $c1 = 0.01^2$, $c2 = 0.03^2$, respectively.

**Table 2**
The detailed structure of the decoder network.

| Layers | Input | Layer type | Filters | Kernel_size | Stride and padding | Output shape |
|---|---|---|---|---|---|---|
| Center | $15 \times 15 \times 512$ | Conv2d + ReLu | 512 | 3 | (1, 1),(1, 1) | $15 \times 15 \times 512$ |
|  | $15 \times 15 \times 512$ | Conv2d + ReLu | 256 | 3 | (1, 1),(1, 1) | $15 \times 15 \times 256$ |
| Block5 | $15 \times 15 \times 768$ | Upsampling | – | – | – | $30 \times 30 \times 768$ |
|  | $30 \times 30 \times 768$ | Conv2d + ReLu | 512 | 3 | (1, 1),(1, 1) | $30 \times 30 \times 512$ |
|  | $30 \times 30 \times 512$ | Conv2d + ReLu | 256 | 3 | (1, 1),(1, 1) | $30 \times 30 \times 256$ |
| Block4 | $30 \times 30 \times 768$ | Upsampling | – | – | – | $60 \times 60 \times 768$ |
|  | $60 \times 60 \times 768$ | Conv2d + ReLu | 512 | 3 | (1, 1),(1, 1) | $60 \times 60 \times 256$ |
|  | $60 \times 60 \times 256$ | Conv2d + ReLu | 256 | 3 | (1, 1),(1, 1) | $60 \times 60 \times 256$ |
| Block3 | $60 \times 60 \times 512$ | Upsampling | – | – | – | $120 \times 120 \times 512$ |
|  | $120 \times 120 \times 512$ | Conv2d + ReLu | 256 | 3 | (1, 1),(1, 1) | $120 \times 120 \times 256$ |
|  | $120 \times 120 \times 64$ | Conv2d + ReLu | 64 | 3 | (1, 1),(1, 1) | $240 \times 240 \times 256$ |
| Block2 | $240 \times 240 \times 192$ | Upsampling | – | – | – | $480 \times 480 \times 192$ |
|  | $480 \times 480 \times 192$ | Conv2d + ReLu | 128 | 3 | (1, 1),(1, 1) | $480 \times 480 \times 128$ |
|  | $480 \times 480 \times 128$ | Conv2d + ReLu | 32 | 3 | (1, 1),(1, 1) | $480 \times 480 \times 32$ |
| Block1 | $480 \times 480 \times 96$ | Conv2d | 32 | 1 | (1, 1),(1, 1) | $480 \times 480 \times 32$ |
| Final | $480 \times 480 \times 32$ | Conv2d | 3 | 1 | (1, 1),(1, 1) | $480 \times 480 \times 3$ |

**Table 3**
The detailed structure of the classifier network. The output of classifier4 depends on the number of classes.

| Layers | Layer type | Input features | Output features | Bias |
|---|---|---|---|---|
| Classifier1 | Linear | 512 | 256 | True |
| Classifier2 | Linear | 256 | 128 | True |
| Classifier3 | Linear | 128 | 64 | True |
| Classifier4 | Linear | 64 | No. of classes | True |

For the quality labeling task, we used the classification loss function $L_c$, cross-entropy (CE), which depends on the predicted class from the classifier $\hat{y}$ and corresponding target value $y$. $CE$ is defined as follows:

$$L_c(\hat{y}_i, y_i) = -\sum_{i=1}^{n} y_i \cdot log(\hat{y}_i), \quad (4)$$

where $\hat{y}_i$ is the $i$th scalar value in the model output, $y_i$ is the corresponding target value, and the output size is the number of scalar values in the model output. This loss is an excellent measure of how distinguishable two discrete probability distributions are from each other. In this context, $y_i$ is the probability that event $i$ occurs, and the sum of all $y_i$ is 1, meaning that precisely one event may occur. The minus sign ensures the loss gets smaller when the distributions get closer.

The final objective loss function, $L$, to optimize the FGR-Net model, including the autoencoder and classifier networks, is the combination between the reconstruction loss $L_{rec}$ and the classification loss function $L_c$, as:

$$L = \alpha L_{rec}(\hat{A}, A) + (1 - \alpha) L_c(\hat{y}_i, y_i), \quad (5)$$

where $\alpha$ is a weight factor set to 0.5 in this work.

Experimentally, the best objective loss function for training the whole model is the cross-entropy for the classifier with MSE as a reconstruction loss function.

## 4. Interpretability of deep learning models

Fig. 4 represents the summary of the applied explainability process of FGR-Net in the testing phase. The first part represents the encoder network, the second part represents the classification network, and the third part the module for interpreting the results of the classifier using the Gradient, GradCAM, and Occlusion algorithms, respectively.

We briefly describe three well-known feature attribution methods: Gradient, GradCAM and Occlusion. Fig. 5 shows the result of each of these methods on a typical fundus image. For the occlusion method, green values (positive) indicate regions where the occlusion increases the class score and vice versa for red values. For the Gradient method, green values indicate that increasing the pixel's brightness causes the score to increase. For the GradCAM, the interpretation is more complex since it depends on the representation of the input by the encoder. Therefore, the absolute value of GradCAM's output is usually interpreted as the importance of each region for a specific class regardless of sign.

*Gradient.* The simplest feature attribution method computes the Gradient of an output score concerning the input. In the case of images, let $f : R^{H \times W \times C} \implies R$ be the function that represents a neural network and let $L$ be a loss or score function. Then we can compute the Gradient of $L$ concerning an input image $x$ simply by using back-propagation, obtaining:

$$L'(x) = \frac{\partial L(f(x))}{\partial x}, \quad L' : R^{H \times W \times C} \to R^{H \times W \times C}. \quad (6)$$

Therefore $L'$ can indicate how to change an input image to optimize for a score locally. A typical choice for $L$ can be the score for a particular class so that $L'$ indicates how to modify $x$ to improve the score for that class.

While simplest to compute and interpret, the Gradient method suffers from some disadvantages, namely saturation of the activation functions, gradient discontinuities, and thresholding artifacts (Shrikumar, Greenside, & Kundaje, 2017).

*GradCAM.* GradCAM follows the Gradient's method basic idea but adds a projection step so that the gradient can be computed concerning any intermediate layer and then projected back into the input to obtain importance scores (Selvaraju et al., 2019). GradCAM been widely used to analyze deep learning models' representations. Since most researchers have used this method to understand fundus image analysis models, we briefly describe GradCAM's algorithm.

Let $f : R^{H \times W \times C} \implies R$ be a network composed of a feature network $g$ that outputs a feature map and an output network $h$ such that:

$$f(x) = h(g(x)), \quad (7)$$

where $x$ is the input to the whole network. Let $L : R^n \implies R$ be a loss function for $f$. GradCAM computes a saliency score $s(x) : R^{H \times W}$ for each pixel in the input $x$. Note that this score is a function of the input image $x$, so different images produce a different GradCAM score.

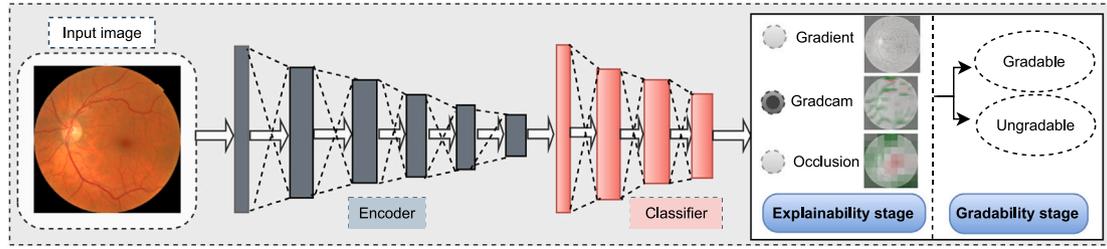

**Fig. 4.** The explainability process for our model in the testing phase representing the results with the Gradient, GradCAM and Occlusion algorithms, respectively.

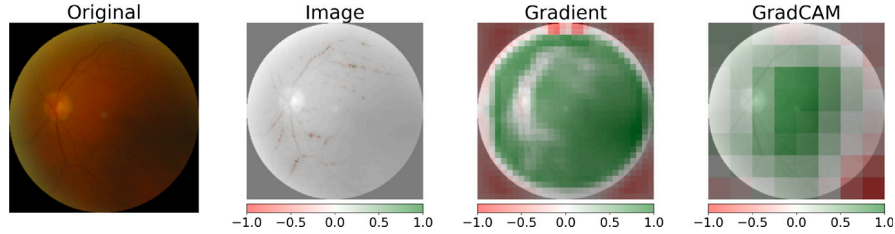

**Fig. 5.** Demonstration of three interpretability methods for a fundus image. From left to right: original image, Gradient, GradCAM and Occlusion.

To obtain $s(x)$, GradCAM calculates the gradients of the loss $L(f(x))$ with respect to the feature maps in the representation:

$$L_g(x) = \frac{\partial L(h(g(x)))}{\partial g(x)}, \quad L_g(x) : R^{H \times W \times C} \to R^{H' \times W' \times C'} \quad (8)$$

Afterwards, GradCAM computes the average of this gradient over the channel dimension to obtain an average value of the importance of each pixel in the feature space, obtaining a tensor:

$$L_g^c(x) = Mean_c L_g(x), \quad L_g^c(x) : R^{H \times W \times C} \to R^{H' \times W'}. \quad (9)$$

Finally, since we are interested in understanding the importance of each pixel or region in the input space, the tensor is resized via bilinear sampling (projected) to match the input image $x$, so that:

$$s(x) = Bilinear(L_g^c(x), H \times W), \quad s(x) : R^{H \times W \times C} \to R^{H \times W}, \quad (10)$$

where s(x) gives a score of importance for each input pixel; this last step assumes translational same-equivariance of the features $g$ for the method to work correctly.

Optionally, the scores are transformed via a $ReLU$ function to retain only inputs that positively contribute to the class score.

*Occlusion methods.* Occlusion methods work by replacing subsets of the inputs with a baseline or reference value and analyzing the difference in the output score of the network for specific inputs. In this fashion, they are similar to the Gradient method but less local since the changes or perturbations of the input are much stronger than the infinitesimals used for computing the gradient. For images, occlusion analysis is typically performed by systematically replacing rectangular-sized regions of the image with gray pixels to simulate occlusion of an input feature (Zeiler & Fergus, 2014).

In this work, we tested the three aforementioned explainability methods with the FGR-Net model combining the autoencoder and with classifier networks and the standard method based on a classifier network to compare and explain the attributes of each model. The results will be detailed in Section 5.7.

## 5. Experimental results

This section introduces the datasets, evaluation metrics, and experiments performed to evaluate the FGR-Net model and the interpretation of the FGR-Net features.

### 5.1. Datasets

Three fundus image gradability datasets were used in our experiments: EyePACS (Foundation, 2019), Eye-Quality (EyeQ) (Fu et al., 2019), and an in-house dataset collected in the Hospital Universitari Sant Joan de Reus (HUSJR), Spain. Below, we briefly describe each dataset:

- EyePACS is a publicly available dataset containing 31,031 fundus images categorized into two classes: gradable and ungradable. EyePACS is divided into a training set of 29,033 images and a test set of 1999.
- EyeQ has 28,792 fundus images derived from the EyePACS dataset. Unlike EyePACS, EyeQ categorizes fundus images into three gradability classes: Good, Usable, and Reject. EyeQ is divided into a training set of 12,543 images and a test set of $16,249 images$.
- The in-house dataset was collected from HUSJR. It categorizes fundus images into two gradability classes: gradable and ungradable. In our experiments, we used this dataset as a test set of 1127 images.

### 5.2. Data augmentation

In this work, we applied different data augmentation techniques suggested in Fu et al. (2019) and Foundation (2019) to increase the number of training samples by applying different transformations to each fundus image to diversify the training data, such as random rotations and flipping. For EyePACS, the dataset consists of two classes: (1) gradable with 21,812 images and (0) ungradable classes containing 7218 images. To construct a balanced dataset, after argumentation, class 0 has 28,872 images, and class 1 has 29,083 images. Total training data (i.e. gradable and ungradable) has 57,957 images. In turn, the EyeQ dataset, after augmentation, has 65,876 retinal images balanced split into a three-level quality grading system ('Good', 'Usable', and 'Reject'). Fig. 10 shows some examples of the transformations applied to each input image (see Fig. 6).

### 5.3. Parameter settings

We used the Adam optimizer (Kingma & Ba, 2014) with $\gamma = 0.1$ and an initial learning rate of 0.001. A batch size of 2 and 50 epochs yielded the best combination. All experiments ran on a 64-bit Core I7-6700, 3.40 GHz CPU with 8 GB of memory, and an NVIDIA GTX 1080 GPU on Ubuntu 16.04. We used the PyTorch deep learning framework (Paszke, Gross, Chintala, & Chanan, 2017) for the model implementation. The computational time of FGR-Net for the training process takes around 1 hour and 13 minutes, 37 seconds for each epoch with a batch size of 2. The training speed of our model is around 0.076 per second for each image, and the inference speed is around 0.026 per second for each image.

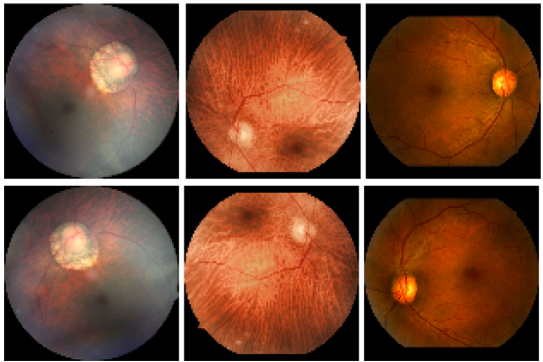

**Fig. 6.** Examples of transformations (flipping and rotation) operations applied to an input image of each class of the EyeQ dataset.

### 5.4. Evaluation metrics

In this work, we used four metrics to evaluate the performance of classification fundamentals as an image gradability-based deep learning model: Accuracy, Precision, Recall, and F1 score.

*Accuracy* that calculates the number of correct predictions divided by the number of predictions. *Precision* refers to the ratio of true positives to the sum of false positives and true positives. The third measure is *Recall* referred to as sensitivity, and defined as the ratio of true positives to the sum of positives. In turn, the fourth measure represents the *F1 score* as the weighted harmonic average of accuracy and recall (Sokolova, Japkowicz, & Szpakowicz, 2006), and it is considered one of the most critical performance measures of the classification model used for medical applications. The closer the value of the F1 score to 1.0, the better the expected performance of the model.

$$Accuracy = \frac{(TP + TN)}{(TP + TN + FP + FN)}, \quad (11)$$

$$Precision = \frac{TP}{(TP + FP)}, \quad (12)$$

$$Recall(Sensitivity) = \frac{TP}{(TP + FN)}, \quad (13)$$

$$F1 Score = \frac{2TP}{(2TP + FP + FN)}, \quad (14)$$

where TP is the number of the true positive samples, TN is the number of the true negative samples, FP is the number of the false positive samples and FN is the number of the false negative samples.

### 5.5. Ablation study

In this section, we study the performance of various CNNs as backbone for the autoencoder network of the FGR-Net model. Particularly, we employed CoAtNets (Dai, Liu, Le and Tan, 2021), ResNet50, Resnet101, Resnet152 (He, Zhang, Ren, & Sun, 2016), wide-resnet (Zagoruyko & Komodakis, 2016), DenseNet121, DenseNet201, DenseNet169 (Huang, Liu, Van Der Maaten, & Weinberger, 2017), SE-Net154, SE-ResNet-101, SE-ResNet-152, SE-ResNet-50, SE-ResNeXT101 (Hu, Shen, & Sun, 2018), and VGG16 (Simonyan & Zisserman, 2014).

For each backbone network, we trained the FGR-Net model two times; the first with two classes using the EyePACS dataset (Good, Reject), and the second with three classes with the Eye-Quality (EyeQ) dataset (Good, Usable, Reject). All models were trained with MSE as a reconstruction loss function (i.e., MSE yields the best results).

The quantitative results are shown in Table 4. Regarding the two-class gradability problem, VGG16 provided the best results among all tested backbones with an Accuracy of 0.8958, Recall of 0.8953, and F1 score of 0.8955. However, the DenseNet169 network achieved the best Precision value (0.8964) among all evaluated backbones. In general, the backbones of ResNet101, DenseNet121, DenseNet169, DenseNet201 and VGG16 achieve similar results in terms of the four measures with values of around 0.89.

In turn, as shown in Table 4, for the three-class problem, we repeated the process with 12 backbones. ResNet152 and VGG16 provided an accuracy of ≥0.89 outperforming the other backbones. SE-ResNet-50 and VGG16 achieved the best results among all tested backbones with a Precision of ≥0.88. ResNet152, DenseNet169, SE-ResNet-50 and VGG16 provided Recall and F1 score of ≥0.87, outperforming the other backbones.

In conclusion, for two-class and three-class problems, it is evident that the VGG16 backbone yields the best results in terms of the four metrics. Thus we select VGG16 as an encoder network for our autoencoder, as shown in Fig. 3.

In order to select the best reconstruction loss function, we trained the autoencoder based on the VGG16 backbone with three loss functions as reconstruction loss to choose the best loss function that can yield the best results. Along with the $MSE$ loss function, we tested two other loss functions: $MAE$ and $SSIM$ functions, explained in Section 3.2. Our experiments with the three loss functions are shown in Fig. 7. With the EyepACS dataset, the MSE loss function outperformed the other functions (i.e., MAE and SSIM) in terms of the four evaluation metrics (i.e., Accuracy, Recall Precision and F1-score). In turn, with the EyeQ dataset, FGR-Net with $MSE$ also yields the best accuracy, recall and F1-score results compared to the two other loss functions (i.e., MAE and SSIM). While FGR-Net with $SSIM$ yields the best results among the three loss functions in terms of precision. Thus, we select the $MSE$ loss function to optimize the autoencoder network to minimize the error between the reconstructed and input images.

### 5.6. Comparisons with state of the arts

#### 5.6.1. Two-class model evaluation

To our knowledge, there are no deep learning-based models related to the two-class problem (gradable and ungradable). Most proposals focused on the level of image quality consisting of a three-class problem (good, usable and reject). Thus, in order to evaluate our two-class FGR-Net model merging the autoencoder and classifier networks, we updated the MCF-Net model proposed in Fu et al. (2019) to work with two classes of gradability. Thus, we retrained the MCF-Net model from scratch with the EyePACS dataset by changing the last layer's structure to work with two classification levels. As shown in Table 5, we find the FGR-Net model based on VGG16 as an encoder network with $MSE$ as a reconstruction loss function outperforms the adapted MCF-Net model with two classes and the two variations of FGR-Net with $MAE$ and $SSIM$ loss functions in terms of the four measures. For instance, the F1-score with FGR-Net yielded a significant improvement of 6% more than the adapted MCF-Net model. Besides, our model combining the autoencoder and classifier networks achieved an accuracy of ≥89% with an improvement of 8% compared to the adapted MCF-Net model. A significant improvement of 25% appeared with the Precision measure compared to the MCF-Net. In addition, the FGR-Net model with the two other loss functions outperforms the adapted MCF-Net in terms of the four measures (Accuracy, Precision, Recall and F1-score) with significant improvements of 7%, 24%, 4%, and 5%, respectively.

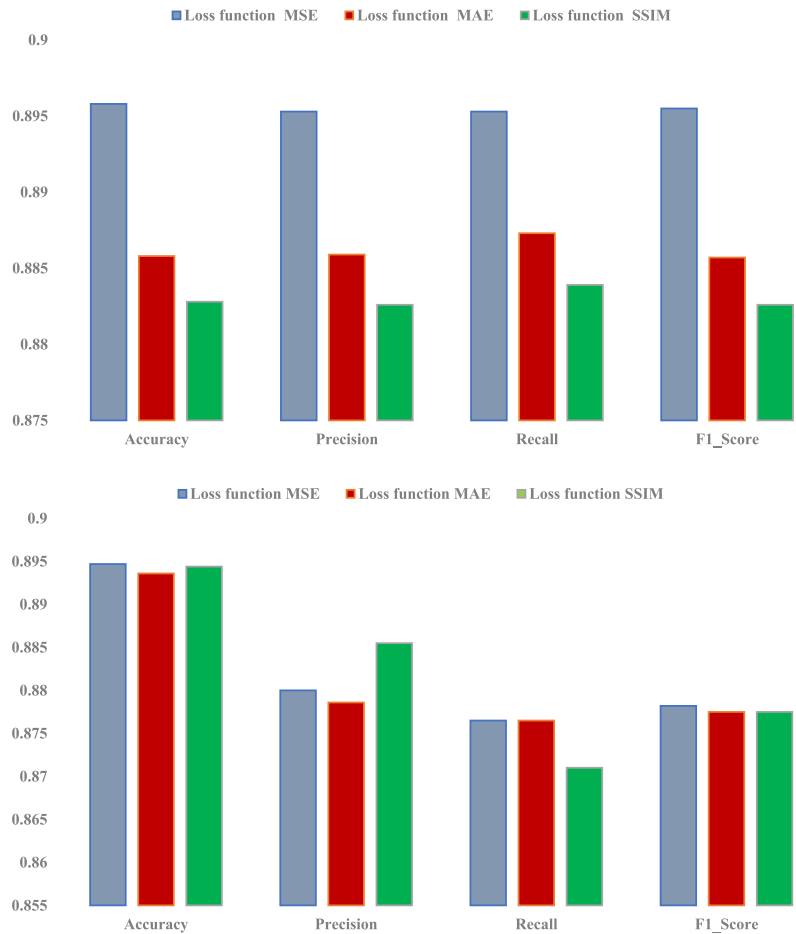

**Fig. 7.** Comparison of the FGR-Net model with three loss functions MSE, MAE and SSIM with (left) the EyePACS dataset, and (right) the EyeQ dataset.

**Table 4**
Evaluation of the FGR-Net model based on different backbones for the autoencoder network on both EyePACS and Eye-Quality (EyeQ) datasets.

|   | Backbones | Accuracy | Precision | Recall | F1 score |
|---|---|---|---|---|---|
| The EyePACS dataset with 2 classes | Autoencoder_CoatNet_0 | 0.8863 | 0.8875 | 0.8885 | 0.8862 |
|  | Autoencoder_Resnet50 | 0.8702 | 0.8711 | 0.7822 | 0.8701 |
|  | Autoencoder_Resnet101 | 0.8938 | 0.8934 | 0.8947 | 0.8936 |
|  | Autoencoder_Resnet152 | 0.8848 | 0.8857 | 0.8868 | 0.8847 |
|  | Autoencoder_Wide_resnet50_2 | 0.8753 | 0.8799 | 0.8789 | 0.8752 |
|  | Autoencoder_Dencenet121 | 0.8918 | 0.8921 | 0.8934 | 0.8917 |
|  | Autoencoder_Dencenet201 | 0.8903 | 0.8911 | 0.8923 | 0.8902 |
|  | Autoencoder_Dencenet169 | 0.8903 | 0.8964 | 0.8945 | 0.8952 |
|  | Autoencoder_SE-ResNet101 | 0.8853 | 0.8855 | 0.8868 | 0.8852 |
|  | Autoencoder_SE-ResNet152 | 0.8873 | 0.8870 | 0.8867 | 0.8868 |
|  | Autoencoder_SE-ReNet-50 | 0.8888 | 0.8887 | 0.8901 | 0.8887 |
|  | Autoencoder_SE_ResNeXT-101 | 0.8853 | 0.8889 | 0.8886 | 0.8853 |
|  | Autoencoder_SE-Net154 | 0.8858 | 0.8852 | 0.8859 | 0.8855 |
|  | ***Autoencoder_Vgg16 (FGR-Net)*** | ***0.8958*** | ***0.8953*** | ***0.8958*** | ***0.8955*** |
| The Eye-Quality (EyeQ) dataset, with 3 classes | Autoencoder_CoatNet_0 | 0.8642 | 0.8426 | 0.8654 | 0.8521 |
|  | Autoencoder_Resnet50 | 0.8892 | 0.8703 | 0.8654 | 0.8676 |
|  | Autoencoder_Resnet101 | 0.8868 | 0.8707 | 0.8676 | 0.8690 |
|  | Autoencoder_Resnet152 | 0.8900 | 0.8754 | 0.8757 | 0.8755 |
|  | Autoencoder_Wide_resnet50_2 | 0.8577 | 0.8357 | 0.8421 | 0.8380 |
|  | Autoencoder_Dencenet121 | 0.8894 | 0.8803 | 0.8687 | 0.8735 |
|  | Autoencoder_Dencenet201 | 0.8866 | 0.8736 | 0.8748 | 0.8736 |
|  | Autoencoder_Dencenet169 | 0.8881 | 0.8696 | 0.8752 | 0.8723 |
|  | Autoencoder_SE-ResNet101 | 0.8877 | 0.8703 | 0.8744 | 0.8720 |
|  | Autoencoder_SE-ResNet152 | 0.8882 | 0.8811 | 0.8599 | 0.8694 |
|  | Autoencoder_SE-ReNet-50 | 0.8929 | 0.8734 | 0.8794 | 0.8751 |
|  | Autoencoder_SE_ResNeXT-101 | 0.8913 | 0.8766 | 0.8723 | 0.8742 |
|  | Autoencoder_SE-Net154 | 0.8707 | 0.8483 | 0.8524 | 0.8479 |
|  | ***Autoencoder_Vgg16 (FGR-Net)*** | ***0.8947*** | ***0.8800*** | ***0.8765*** | ***0.8782*** |

**Table 5**
Comparison between the FGR-Net model and MCF-Net (Fu et al., 2019) on the Eypces (Foundation, 2019) dataset.

| Model | Accuracy | Precision | Recall | F1 score |
|---|---|---|---|---|
| MCF_Net (Fu et al., 2019) | 0.8168 | 0.6475 | 0.8494 | 0.8387 |
| FGR-Net_MAE* | 0.8858 | 0.8859 | 0.8873 | 0.8857 |
| FGR-Net_SSIM* | 0.8828 | 0.8826 | 0.8839 | 0.8826 |
| **FGR-Net_MSE*** | *0.8958* | *0.8953* | *0.8958* | *0.8955* |

(A)

| Class | Ungradable | Gradable |
|---|---|---|
| Ungradable | 841 | 98 |
| Gradable | 110 | 947 |

MSE

| Class | Ungradable | Gradable |
|---|---|---|
| Ungradable | 857 | 82 |
| Gradable | 146 | 911 |

MAE

| Class | Ungradable | Gradable |
|---|---|---|
| Ungradable | 848 | 91 |
| Gradable | 143 | 914 |

SSIM

(B)

| Class | Ungradable | Gradable |
|---|---|---|
| Ungradable | 302 | 72 |
| Gradable | 171 | 582 |

MSE

**Fig. 8.** Confusion matrices with the testing sets of the Eyepaces (A) and Hospital Universities Sant Joan de Reus (B) datasets with FGR-Net and with three different loss functions: $MSE$, $MAE$ and $SSIM$, respectively.

To validate the performance of the FGR-Net model, we computed the confusion matrix and the overall classification accuracy on the test set of the EyePACS dataset and in-house dataset from HUSJR, with the two-class problem (gradable, ungradable). The confusion matrix allows us to have a more detailed analysis than the classification rate. Fig. 10 shows the confusion matrix resulting from the three loss functions with the two datasets. As shown in Fig. 10, the true positives (TPs) and true negative (TNs) of FGR-Net with the EPACS dataset are 947, 843, respectively, out of 1996 images with the $MSE$ loss function. The $MAE$ yielded TPs and TNs of 911 and 857, respectively. In turn, TPs and TNs are 914, and 848, respectively, with the $SSIM$ loss function. In the case of the in-house dataset, we are using the best loss function $MSE$. For evaluation, the TPs and TNs obtained by the FGR-Net model are 582 and 302, respectively, out of 1127 images as a test set (see Fig. 8).

*5.6.2. Three-class evaluation*

In order to compare the FGR-Net model with modern gradability assessment methods, we compared FGR-Net to the state-of-art of three-class (good, usable, Reject) gradability assessment on the public Eye-Quality (EyeQ) dataset. Table 6 summarizes the results of the three variations of FGR-Net with eight methods: two methods are based on hand-crafted features; BRISQUE (Mittal, Moorthy, & Bovik, 2012) and NBIQA (Ou, Wang, & Zhu, 2019), and six methods based on deep learning designed for retinal fundus image quality assessment; TS-CNN (Yan, Gong, & Zhang, 2018), HVS-based method (Wang et al., 2015), MCF-Net (Fu et al., 2019), multivariate regression CNN (MR-CNN) (Raj et al., 2020), the Double branch network, SalStructIQA (Xu, Liu et al., 2020) and multi-level quality assessment network (Muddamsetty & Moeslund, 2020). A three variations, with the three loss functions, of FGR-Net combining the autoencoder with classifier networks provided the best results in terms of Accuracy, Precision, Recall and F1-score. Among the three variations, the model with MSE as a loss function yields the best results in terms of Accuracy, Recall and F1-score. Thus, we showed only the results of FGR-Net with MSE as a loss function in Table 6. In turn, The model with MAE achieved the best Precision value. As shown in Table 6, our model significantly improved the F1-score by 14% compared to the handcrafted methods. Our model achieved a small improvement with the four measures compared to SalStructIQA (Xu, Liu et al., 2020) and CNN-combined (Muddamsetty & Moeslund, 2020). However, the method proposed in Muddamsetty and Moeslund (2020) combines deep and handcrafted features for assessing the fundus image quality. In turn, the method proposed in Xu, Liu et al. (2020) segments two salient features before classifying the quality of fundus images that adds more complexity to their model. In contrast, FGR-Net is a straightforward model. Since, we only used the encoder network and the classifier in the testing stage without extracting prior information from the input fundus images.

In order to check the scalability and upgradability of the FGR-Net model on the EyeQ dataset with three-classes ((0) Good, (1) Usable, (2) Rejected), we also computed the confusion matrix with three loss functions (MSE, MAE, SSIM) and overall classification accuracy in the test set. As explained above in the subsection, this allows for a more detailed analysis than just a high rating ratio. As shown in 9, TPs and TNs of FGR-Net with a test set of 16,255 images with three loss functions. The model was able to classify the fundus images into three classes with a small number of mispredictions. For instance, the model with MSE and the first class "Good" classified only five "Reject" images as good images and 387 "Usable" images as "Good" images. This result is intuitive since both "Usable" and "Good" images have similar characteristics. Among of the three reconstruction losses, MSE yields the highest TP and TN on the test set of the EyeQ dataset. FGR-Net achieved an AUC of 0.94 with class 0, 0.88 with class 1, and 0.91 with class 2. The results confirm the results shown in the confusion matrix, see the supplementary materials for more results.

*5.7. Limitations and robustness*

For the image gradability classification into three classes (i.e., Good, Usable and Reject), our model, FGR-Net, has one limitation related to the Usable class. Due to the underrepresentation in the data, the model may struggle to learn its characteristics and classify them accurately in some cases. To assess that, we have included Fig. 10, which shows 4 misclassified samples. These qualitative results support the quantitative results of the confusion matrix shown in Fig. 9. However, the good thing is that most misclassified Usable images are recognized as Good. This is acceptable from the clinical point of view since ophthalmologists can use Usable images for the diagnosis and screening, like Good images. Additionally, the Usable images are sometimes unclear to ophthalmologists; thus, the model may classify them as a Reject class. Thus, the FGR-Net model mimics human beings with the images of the Usable class.

To assess the robustness of the FGR-Net model and to show how the model behaves when presented with noisy or perturbed images, we systematically tested our model with six types of common perturbation methods:

1. Gaussian Blur ($\sigma = [0.5, 1.5]$) (Gedraite & Hadad, 2011)
2. Additive Gaussian Noise (scale = [0.5, 0.04*255]) (Bergmans, 1974)
3. Gamma Contrast (scale = [0.5, 1.5]) (Huang, Cheng, & Chiu, 2012)
4. Additive Poisson Noise ($\lambda = 10.0$) (Rodrigues, Sanches, & Bioucas-Dias, 2008)
5. Affine Transformation (scale = [0.5, 1.5]) (Weisstein, 2004)

Table 6
Comparison of the FGR-Net model with different methods of the existing methods with the Eye-Quality (EyeQ) dataset (Fu et al., 2019).

| Method | Accuracy | Precision | Recall | F1 score |
|---|---|---|---|---|
| BRISQUE (Mittal et al., 2012) | 0.7692 | 0.7608 | 0.7095 | 0.7112 |
| NBIQA (Ou et al., 2019) | 0.7917 | 0.7641 | 0.7509 | 0.7441 |
| TS-CNN (Yan et al., 2018) | 0.7926 | 0.7976 | 0.7446 | 0.7481 |
| HVS-based (Wang et al., 2015) | – | 0.7404 | 0.6945 | 0.6991 |
| MR-CNN (Raj et al., 2020) | 0.8843 | 0.8697 | 0.8700 | 0.8694 |
| DenseNet121-MCF (Fu et al., 2019) | – | 0.8645 | 0.8497 | 0.8551 |
| DenseNet121-MCF (Fu et al., 2019) | 0.8722 | 0.8563 | 0.8482 | 0.8506 |
| DenseNet121-RGB (Fu et al., 2019) | – | 0.8194 | 0.8114 | 0.815 |
| DenseNet121-RGB (Fu et al., 2019) | 0.8568 | 0.8481 | 0.8239 | 0.8315 |
| ResNet-18-RGB (Fu et al., 2019) | – | 0.804 | 0.816 | 0.808 |
| ResNet-18-HSVB (Fu et al., 2019) | – | 0.801 | 0.816 | 0.808 |
| ResNet-50-RGBB (Fu et al., 2019) | – | 0.812 | 0.807 | 0.810 |
| Resenet-50-HSVB (Fu et al., 2019) | – | 0.770 | 0.777 | 0.773 |
| Single-branch SalStructIQA (Xu, Liu et al., 2020) | 0.8847 | 0.8715 | 0.8645 | 0.8662 |
| Dual-branch SalStructIQA (Xu, Liu et al., 2020) | 0.8897 | 0.8748 | 0.8721 | 0.8723 |
| CNN-RGB (Muddamsetty & Moeslund, 2020) | – | 0.860 | 0.862 | 0.860 |
| CNN combined (Muddamsetty & Moeslund, 2020) | – | 0.878 | 0.880 | 0.878 |
| **FGR-Net** | **0.8947** | **0.8800** | **0.8765** | **0.8782** |

(A)

| Class | Good | Usable | Reject |
|---|---|---|---|
| Good | 8079 | 387 | 5 |
| Usable | 533 | 3784 | 291 |
| Reject | 19 | 531 | 2670 |

MSE

(B)

| Class | Good | Usable | Reject |
|---|---|---|---|
| Good | 8009 | 450 | 12 |
| Usable | 482 | 3709 | 367 |
| Reject | 21 | 397 | 2802 |

MAE

(C)

| Class | Good | Usable | Reject |
|---|---|---|---|
| Good | 8018 | 441 | 12 |
| Usable | 461 | 3750 | 347 |
| Reject | 16 | 434 | 2770 |

SSIM

**Fig. 9.** Confusion matrices with the testing sets of the Eye-Quality (EyeQ) dataset with the FGR-Net model and with three different loss functions: (A) $MSE$, (B) $MAE$ and (C) $SSIM$, respectively.

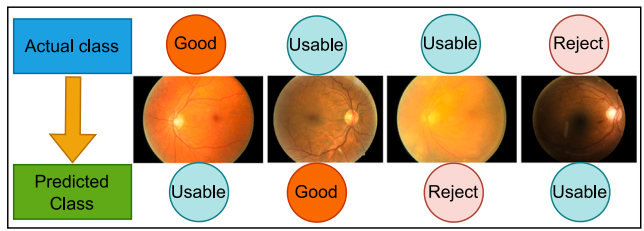

**Fig. 10.** Four misclassified images of the EyeQ dataset, including their true label and that predicted by the model.

6. Multiplicative Noise (scale = [0.1, 5.5]) (Sancho, San Miguel, Katz, & Gunton, 1982)

We applied the six perturbation methods on 100 images selected from the test set. Table 7 shows the evaluation of our model with the six perturbations. The model achieved satisfactory results with all perturbations, especially Multiplicative Noise and Gamma Contrast, with a reduction of 1% to 2%. The model's performance was reduced by 5% with the Adaptive Gaussian. This type of noise can be particularly difficult for FGR-Net and deep learning models in general to handle because it can introduce significant changes in the data that are not easily discernible to the human eye. Also, our model is designed to learn complex visual patterns in the fundus images, and these patterns may not hold up well when the Adaptive Gaussian noise is introduced. The Adaptive Gaussian. However, we can cope with this noise by generating noisy images with Adaptive Gaussian via an augmentation process and training the model by such images. Thus, based on the results, we can say that The FGR-Net is a robust model for different types of noise and perturbations.

Table 7
Evaluation of our model with six different perturbation methods.

| Noise type | Accuracy | Precision | Recall | F1-score |
|---|---|---|---|---|
| FGR-Net | **0.8966** | **0.8981** | **0.8966** | **0.8956** |
| GaussianBlur | 0.8400 | 0.8556 | 0.8400 | 0.8423 |
| AdditiveGaussian | 0.8266 | 0.8321 | 0.8266 | 0.8268 |
| GammaContrast | 0.8733 | 0.8742 | 0.8733 | 0.8737 |
| AdditivePoisson | 0.8400 | 0.8500 | 0.8400 | 0.8394 |
| Affine | 0.8300 | 0.8364 | 0.8300 | 0.8302 |
| Multiply | 0.8866 | 0.8888 | 0.8866 | 0.8850 |

*5.8. Interpretation of model features*

We used various interpretability methods to understand the Normal model based on a traditional classification network and our model, which combines an autoencoder and a classifier, to compare their internal representations. We focus on the two-class problem (gradable vs ungradable) since we are interested in determining which features the model uses to determine whether a fundus image is of good quality. Our approach employs:

1. Saliency map methods such as Gradient and GradCAM visualizations to understand the relevance of the input regions (Escalante et al., 2018).
2. Unsupervised learning methods to identify common patterns in activations (Kolouri, Martin, & Hoffmann, 2017).
3. Generative and Adversarial techniques to understand model robustness and class definitions (Madry, Makelov, Schmidt, Tsipras, & Vladu, 2017).

Our experiments use PyTorch models via the Captum interpretability package (Kokhlikyan et al., 2020).

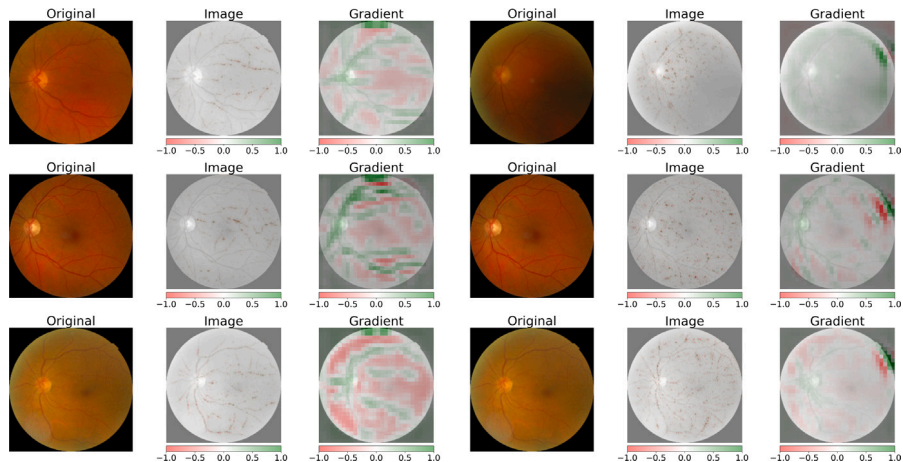

**Fig. 11.** Comparison of the saliency and GradCAM visualizations for three samples of class *gradable*, between the two models. The left group corresponds to the autoencoder-based model, while the right group corresponds to the Normal model.

#### 5.8.1. Saliency maps

We measured the importance of the input regions for the gradability classification task via Saliency maps. These allow us to understand which features the model *expects* to perform and assign a class to an input image.

We use two saliency map methods that yield different insights into the model. First, the classical Gradient method computes the output gradient concerning the input pixels (Escalante et al., 2018), focusing more on border-like information such as blood vessels (Molnar, 2019). Second, we employed GradCAM (Selvaraju et al., 2019) to understand the focus of the representations in the last feature map of the encoder part of the model, given that GradCAM provides a more holistic set of regions of interest (Selvaraju et al., 2019). We do not use Occlusion methods since these are very computationally intensive (see Table 8).

Fig. 11 shows a few input images and the corresponding saliency maps obtained with Gradient and GradCAM, for both models and with samples of the *Gradable* class. The Gradient method focuses on smaller blood vessels of the fundus image for both models. This is expected since this method typically focuses on low-level features (Escalante et al., 2018). However, in the FGR-Net case, the saliency map better highlights these blood vessels more consistently.

On the other hand, GradCAM focuses on more medically relevant structures, such as the optic disk, as well as the main blood vessel region. However, in the case of the Normal model based on the classification network, there is an unexpected and hard-to-explain focus on the upper-right border of the fundus disk, which is not a medically relevant region. Fig. 12 shows the model focuses on the same border highlighted in Fig. 11 but now for samples of the *Ungradable* class. Again, the Normal model focuses on more irrelevant areas, especially the top-right border. On the other hand, the autoencoder model's GradCAM saliency map shows it focuses on relevant eye regions.

While these examples illustrate, they focus on single samples and do not represent the whole complexity of the model's encoding. The supplementary materials (Appendix 2) contain a detailed analysis of saliency maps averaged over a large set of examples and also through unsupervised learning, which supports the previous findings.

#### 5.8.2. Adversarial examples

Adversarial examples can be used to understand the robustness and vulnerability of a model against attacks. In general, attacks are perturbations of the input that are considered intentional and can fool models. However, in this case it would be helpful to consider inadvertent transformations that may perturb the fundus images and thus affect model performance, such as new equipment, calibration, or capture technique.

Fig. 13 shows how two typical fundus images of a class can be transformed into an image of another class via the white-box gradient-based adversarial attack Projected Gradient Descent (PGD) (Madry et al., 2017). We used a targeted PGD with 20 steps, a step size of 0.01, and a radius of 0.13 for all images and targets. In this case, we only focus on the FGR-Net model since both models have similar results with these techniques. Finally, we chose images representative of each class but not ideally classified by the model (around 0.95 model probability). In this way, the images are representative of their class. Still, at the same time, PGD can yield a perturbation that can make a *gradable* image more so in a way that is significant (i.e., can be visualized), and the same for the *ungradable* class.

In all cases, the figure shows the general tendency of the model to prefer backgrounds that are not as black as the inputs (blue values around the borders). This behavior is interesting since the dataset's images had their background segmented out and are, therefore, purely black at the edges, so the data distribution should not be centered on bright backgrounds. This suggests that the encoding of the classes is relative in terms of brightness since the brightness of the fundus is also increased in general. Besides, PGD tends to slightly blur the image while increasing its brightness, washing out the colors, and lowering the contrast. Still, at the same time, it adds additional blood vessels that were non-existent in the original image.

We can see that to transform images into the *Ungradable* class (rows 2 and 4), PGD tends to add small blood vessels, both for the image that was originally *gradable* as for the *Ungradable* case. In turn, to convert images to the *gradable* class, the model through PGD adds wider blood vessels in more specific locations and is more selective concerning smaller blood vessels (rows 1 and 3).

This strategy corresponds with the previous saliency analysis using the Gradient method, indicating that blood vessels are a vital factor in the classification. However, surprisingly, the smaller vessels' location is not essential to the recognition.

On the other hand, the figures show that for *Ungradable* images (rows 3 and 4), PGD almost completely ignores the optic disk region. This is not the case for the *gradable* images (rows 1 and 2), where PGD preserves (white color) the original image, although it produces a slight blurring effect.

Finally, it is also interesting that for the *Ungradable* image, in both cases (rows 3 and 4) the PGD's strategy includes darkening the macula region so that the background's color matches the macula, giving the impression that the fundus mixes with the background. This indicates that features based on blood vessels may be more critical for recognition than those based on brightness. On the other hand, to turn a *gradable* into an *Ungradable* or *gradable* image, the model highlights and slightly decreases the brightness of the optic disk region in both cases. This seems contradictory since a lack of a well-defined optic disk should be a clear sign of an *Ungradable* image and further indicates the focus on the model on the blood vessels.

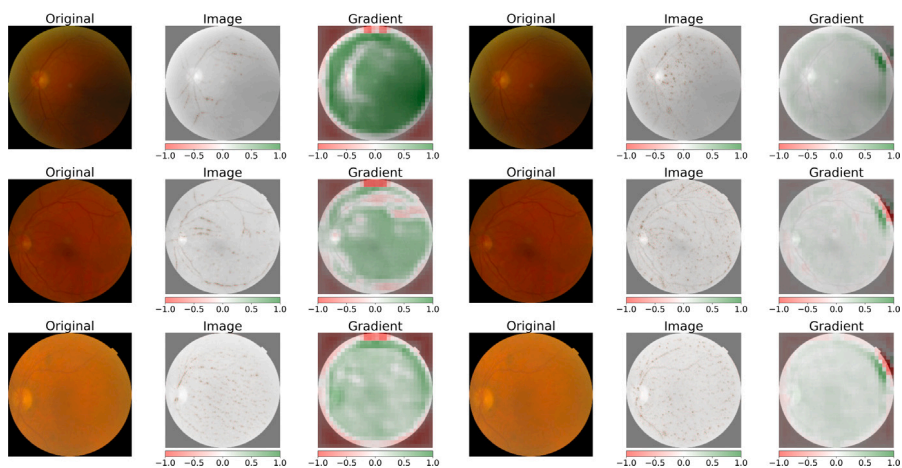

**Fig. 12.** Comparison of the saliency and GradCAM visualizations for three class samples *ungradable*, between the two models. The left group corresponds to the FGR-Net model, while the right group corresponds to the Normal model.

**Table 8**
Average computation time (in milliseconds) over 50 evaluations for different interpretability methods and GPUs.

|  | NVIDIA Geforce 1050ti, 4 GB | NVIDIA Pascal X Titan, 12 GB |
| --- | --- | --- |
| Image prediction | 11 ms | 2 ms |
| Gradient | 115 ms | 26 ms |
| GradCAM | 171 ms | 39 ms |
| Occlusion | 3034 ms | 718 ms |

*5.8.3. Performance evaluation for real-time feedback*

Visualizations of the features on which the model focuses can help medical practitioners and technicians validate the quality in acquiring fundus images. With the advent of mobile fundus photography, real-time feedback on the quality of the fundus images can enable the acquisition of high-quality images with low-cost devices. Therefore, we evaluated the performance of each of the visualization methods on the proposed model. Since the decoder part is not used in the prediction, both the FGR-NET and the Normal model based on a classifier network have the same performance.

We tested both models on GPUs, NVIDIA GeForce 1050TI with 4 GB of RAM and NVIDIA Pascal X Titan with 12 GB of RAM, on an Intel i7 CPU with 16 GB of RAM. We measured the mean computation time over 50 runs. Each run consisted of a batch with a single sample to reflect a real-time setting.

Table 8 shows the computation time for the various methods. The latency for image prediction (row 1) is included as a baseline. In all cases, the coefficient of variation was lower than 0.004 (i.e., unbiased estimator). As Table 8 shows, the Pascal X Titan is capable of real-time performance, achieving 38 fps (26 ms) for the Gradient method and 25 fps (39 ms) for the GradCAM. The 1050ti achieves 9 fps (115 ms) and 6 fps (171 ms) for those methods. Given that these timings were obtained and measured using a stock PyTorch implementation, with single precision and without any optimizations, this performance can be improved with Captum's stock implementation of the interpretability measures.

Additionally, we included the performance of the Occlusion method, whose visualizations were not considered in the analysis because it is very computationally intensive.

## 6. Conclusions

This work proposed a deep learning model, FGR-Net, combining autoencoder and multi-layer classifier networks for predicting the gradability of retinal fundus images. The autoencoder consists of two networks: encoder and decoder. The autoencoder network is used to reconstruct the input fundus image. Our model also includes a multi-layer classifier fed by features extracted from the encoder network to rank the gradability of the fundus image as *gradable* or *ungradable*. FGR-Net's learning approach combines the cross-entropy loss function based on supervised learning and self-supervised learning by comparing the reconstructed image to the target image (i.e., the input image). The FGR-Net model based on the VGG16 backbone as the base of the encoder network and using the MSE as a reconstruction loss function achieved an overall accuracy of 0.8947, precision of 0.8800, recall of 0.8765, and F1-score of 0.8782. Our model outperformed the state-of-the-art retinal gradability assessment in the two-class (gradable and ungradable) and three-class (Good, Usable and Reject) tasks. The FGR-Net model can correctly identify the visual features of eye image gradability for a more precise grading system.

In addition, based on interpretability analysis, we show that the FGR-Net model mainly focuses on the presence and type of blood vessels in the fundus images via the use of three interpretability methods. FGR-Net showed that other vital structures, such as the optic disk and macula, play a lesser role than expected in the gradability of fundus image. The interpretability analysis also found that the addition of the decoder and reconstruction loss helps the FGR-Net model focus more on relevant structures of the fundus image. We also evaluate the computational cost of each interpretability method to determine their ability to run in a real-time context to improve feedback during image acquisition. Our results showed that inexpensive consumer-grade GPUs could provide acceptable performance for real-time computation, even with un-optimized model and interpretability implementations. FGR-Net can generate more interest in the biomedical community to improve the performance of retinal image gradability assessment tasks, which play an essential role in applications such as retinal image segmentation and automatic disease diagnosis.

Future work aims to use the developed assessment model to improve the accuracy of classifying and grading eye diseases (e.g., Diabetic Retinopathy) based on interpretation models.

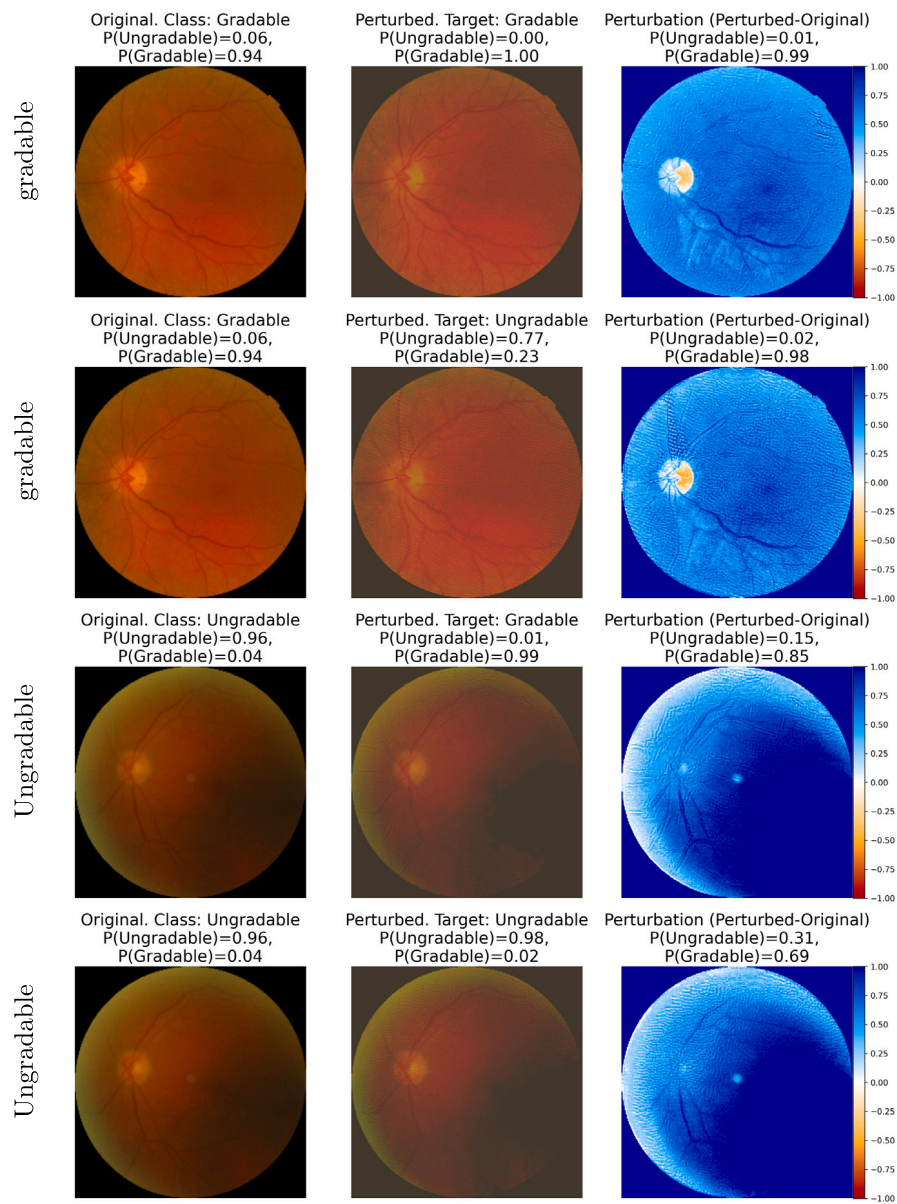

**Fig. 13.** Adversarial perturbations on an input image. The first column shows the original image for a *gradable* sample (rows 1 and 2) and an *Ungradable* sample (rows 3 and 4). The second column shows the perturbed image so that it is closer to samples of class *gradable* (rows 1 and 3) or *Ungradable* (rows 2 and 4). The last column shows the perturbation applied (*Target − Perturbed*), averaged over the three color channels of the input image.

**Declaration of competing interest**

The authors declare that they have no known competing financial interests or personal relationships that could have appeared to influence the work reported in this paper.

**Data availability**

Data will be made available on request.

**Acknowledgment**

This publication has received support from the research project RetinaReadRisk, with EIT Health and Horizon Europe funding under grant agreement 220718.

**Appendix A. Supplementary data**

Supplementary material related to this article can be found online at https://doi.org/10.1016/j.eswa.2023.121644.